\documentclass[cits]{PoS}
\usepackage{amsmath}
\usepackage{amsfonts}
\def\bge{\begin{equation}}
\def\ene{\end{equation}}
\def\bgea{\begin{eqnarray}}
\def\enea{\end{eqnarray}}
\def\nn{\nonumber}

\title{On modeling the scalar meson dynamics with Resonance Chiral Theory}

\ShortTitle{On modeling the scalar meson dynamics with $\rm{R}\chi{T}$}

\author{\speaker{Sergiy IVASHYN}%
         \thanks{Currently at IP, University of Silesia, Katowice PL-40-007, Poland}
 \ \ and Alexandr KORCHIN
\\
 NSC ``Kharkov Institute for Physics and Technology'',
      Institute for Theoretical Physics,\\ Kharkov UA-61108, Ukraine\\
        E-mail: \email{ivashin.s@rambler.ru}, \email{korchin@kipt.kharkov.ua}}

\abstract{
  The features of the Resonance Chiral Theory ($\rm{R}\chi{T}$) related
to the description of the lightest scalar resonances, $\sigma$,
$f_0(980)$ and $a_0(980)$, are discussed. Major attention is paid to the
fits of the invariant mass distributions in the radiative decays of the
$\phi(1020)$ meson ($\phi \to (\gamma a_0 \to) \gamma \pi\eta$ and $\phi
\to (\gamma f_0/\sigma \to) \gamma \pi\pi$). The study of the scalar
sector in $\rm{R}\chi{T}$ is motivated by the success of the theory
predictive power in numerous processes with other types of resonances. We
conclude that  $\rm{R}\chi{T}$ is sufficiently flexible to
describe these decays, however the further quantitative improvement is
required. The technical work-outs and related important questions are
outlined. }


\FullConference{International Workshop on Effective Field Theories: from the pion to the upsilon\\
                 2-6 February 2009\\
                 Valencia, Spain}

\begin{document}

\section{Introduction}

The Resonance Chiral Theory ($\rm{R}\chi{T}$) is a consistent extension of the Chiral Perturbation Theory (ChPT)
to the region of energies near 1 GeV by explicit introduction of the
 resonance fields and exploiting the idea of resonance saturation~\cite{EckerNP321}.
One of the advantages of the $\rm{R}\chi{T}$
Lagrangian at leading order (LO) (which we essentially use in our approach)
is that, having a good predictive power,
it contains very few free parameters compared
with the other phenomenological models.

The fields of $\rm{R}\chi{T}$, which correspond to meson
resonances, are the large-$N_c$ narrow states with equal masses within
the multiplet. The mass splitting corrections were worked-out in a
consistent way for $J^{PC} = 1^{--}, 2^{++}, 1^{++}, 1^{+-}, 0^{-+}$
nonets~\cite{Cirigliano:resonances}. In the light scalar sector ($J^{PC}
= 0^{++}$) the deviation of mass matrix for physical states from its
large-$N_c$ limit is large. There are also other indications that the
%%% $\mathcal{O}(\tfrac{1}{N_c})$
$\mathcal{O}(1/N_c)$ corrections are very important for the
light scalar mesons. For the advances in the understanding the scalar
mesons we refer to the corresponding section in the Particle Data
Review~\cite{PDG_2008} and the Proceedings of the recent Workshop
dedicated to the subject~\cite{SCADRON70}.

% It is not surprising that the inclusion of these particles into
% $\rm{R}\chi{T}$ is not straightforward.
There were numerous attempts to describe the dynamics of
the lightest scalar mesons
--- $a_0(980)$~($I^G = 1^-$), $f_0(980)$ and  $f_0(600)~\equiv~\sigma$~($I^G = 0^+$) ---
by means of chiral theories, e.g.~\cite{Black:1998wt,Ivashyn:2007yy,Harada06}.
In Ref.~\cite{Black:1998wt}, for instance, a decade ago
the light scalars were successfully united into the nonet and the subsequent
symmetry relations were studied.
The consideration of scalar sector in $\rm{R}\chi{T}$
usually avoided the explicit assignments for all of the multiplet members.

The technical work-outs from the $\rm{R}\chi{T}$ for the radiative decays with
the scalar mesons can be found e.g.\! in Ref.~\cite{Ivashyn:2007yy}.
The general features of the approach are sketched in Section~\ref{sec:Framework}.
In Section~\ref{sec:Problems} we pay attention to
the important issues like removal of the scalar tadpole
terms of the  $\rm{R}\chi{T}$ Lagrangian and $\eta$ (and $\eta^\prime$) inclusion in $\rm{R}\chi{T}$.

The estimates of the model parameters in Ref.~\cite{Ivashyn:2007yy}
were carried out in a na\"ive way
(see~\cite{Ivashyn:2008sg}),
% \footnote{We thank E.~Oset and J.A.~Oller for
% bringing our attention for the first time to this peculiarity of scalar meson interactions:
% the parameter values obtained from the integral widths (e.g. from those shown in the PDG review)
% may easily fail to describe the differetial decay properties, e.g. the invariant mass spectra
% in the $\phi$ decays. For illustration see Ref.~\cite{Ivashyn:2008sg}.},
however the interaction pattern and dynamic details remain solid and
easily allow for the other ways of parameter extraction. Improvements on
this way were initiated in Ref.~\cite{Ivashyn:2009yt} and this paper is
aimed to this analysis as well. Section~\ref{sec:Fit} is devoted to the
fits of the invariant mass distributions in the radiative decays of the
$\phi(1020)$ meson: $\phi \to (\gamma a_0 \to) \gamma \pi\eta$ and $\phi
\to (\gamma f_0/\sigma \to) \gamma \pi\pi$. For the extraction of the
mixing parameter and the couplings we try to use the model-independent
information on the pole position of $\sigma$~\cite{Colangelo}, pole
position of $f_0(980)$ due to Ref.~\cite{Pennington-Analysis} and 
information on the $a_0(980)$ parameters from Ref.~\cite{L3}.
Brief conclusions are drawn in Section~\ref{sec:Discussion}.

\section{Scalar mesons in R$\chi$T}
\label{sec:Framework} The scalar resonances below $1$~GeV ($\sigma$,
$f_0(980)$ and $a_0(980)$) have important consequences for the low-energy
hadronic interactions (e.g. they contribute to ChPT LECs in order
$\mathcal{O}(p^4)$). If one does not use any special (non-perturbative)
techniques to account for corrections to the leading order in
$1/N_c$, then to be consistent with the physics one has to
introduce the resulting effects explicitly in the $\rm{R}\chi{T}$
Lagrangian. One may conclude that the large-$N_c$ counting has to be
somewhat relaxed in favor of effects peculiar for the light
scalars,
%%% for that purpose
 at the same time the chiral symmetry has to be
preserved. With the notation of
Refs.~\cite{EckerNP321,Cirigliano:resonances} the relevant Lagrangian
reads
 \bgea \label{lagr:master} \mathcal{L}_{scalar} &=&
\frac{1}{2} \langle \nabla^\lambda S\;
 \nabla_\lambda S - M_S^2 \, S^2 \rangle
+ e^S_m \left\langle S^2\; \chi_+ \right\rangle
+ k^S_m S_0 \left\langle S^{oct} \chi_+ \right\rangle
- \frac{\gamma^S_m \;M_S^2}{2}S_0^2 +\mathcal{L}_{int},
\enea
\bgea
\label{lagr:inter}
\mathcal{L}_{int}&=&
 c_d \left\langle S \; u_\mu u^\mu \right\rangle
+ c_m \left\langle S\; \chi_+ \right\rangle
.
\enea
Scalar octet  $S^{oct}$ and singlet  $S_0$ fields
are related to the physical degrees of freedom as follows:
\bgea \label{eq:multiplet_sc}
\left\{
\begin{aligned}
a_0(980) =& S_3
,\\
f_0(980) =& S_0\, \cos \theta - S_8\, \sin \theta
,\\
\sigma =& S_0\, \sin \theta + S_8\, \cos \theta .
\end{aligned}
\right.
 \enea
Here $S_3$ is the isospin-one, $S_8$ is the isospin-zero
 neutral members of the flavor octet.
%The octet-singlet mixing angle $\theta$ is needed to diagonalize the mass matrix.
There are indications for the
 $\sigma$, $f_0(980)$ and $a_0(980)$ mesons to be members of one
multiplet~\cite{Black:1998wt,Oller:2003vf}.

For the attempts to work out the mass splitting for lightest scalar
mesons from the kinetic and mass part of Lagrangian~(\ref{lagr:master})
we refer to~\cite{Cirigliano:resonances,Black:1998wt}. In the current
paper we assume that the values of $e^S_m$, $k^S_m$ and  $\gamma^S_m$ are
implicitly tuned in such a way, that the mass eigenvalues correspond to the observed poles.
% \footnote{It is known that $e^S_m > 0$ corresponds to the lowest scalar meson mass hierarchy, usually called ``inverted hierarchy''.}.
In principle, the  mixing angle $\theta$ is determined by  the mass
diagonalization, however it also affects the interaction pattern of the
physical states. Within the scope of the paper we fix the $\theta$ from
fit to decay distributions.

In the interaction Lagrangian~(\ref{lagr:inter}) the simplification
$c_{m,d} S = c_{m,d}\left( S^{oct} + S_0/\sqrt{3} \right)$ is assumed. In
notation of Ref.~\cite{EckerNP321} it means that $\tilde{c}_{m,d} =
c_{m,d}/\sqrt{3}$ in the large-$N_c$ limit. The coupling
constants $c_d$ and $c_m$ have to be fixed from the measured decays. We
use the $\phi$ radiative decay distributions for that purpose below in
Section~\ref{sec:Fit}.

\section{Scalar tadpoles, $\eta$ meson and pseudoscalar decay constants in $\rm{R}\chi{T}$}
\label{sec:Problems}

% The Lagrangian~(\ref{lagr:master}) contains the tadpole term~\cite{Meissner:tadpole}
%  for the scalar fields $f_0$ and $\sigma$:
% \bgea
%  \mathcal{L}^{tad}_{S} = {2 c_m} \left\langle S^{} \chi \right\rangle
%  &=&
%  \frac{2 {c}_m}{\sqrt{3}}
%  (
%  (2 m_K^2 + m_\pi^2)\cos \theta
% + 2\sqrt{2}( m_K^2 - m_\pi^2)\sin \theta
%  ) f_0
%  \nn
%  \\
%  &&\hspace{-0.2in}+\,
%  \frac{2 {c}_m}{\sqrt{3}}
%  (
%  -2\sqrt{2}(m_K^2 - m_\pi^2)\cos \theta
% + (2 m_K^2 + m_\pi^2)\sin \theta
%  ) \sigma
% .
% \enea
% It was shown~\cite{JJSC:tadpole} that it is possible
% to consistently eliminate these tadpoles by redefining the scalar fields,
% e.g. $S = \bar{S} + c_m \chi \left[  M_S^2 - 2 e_m^S \chi \right]^{-1}$,
% in~(\ref{lagr:master}).

The pseudoscalar meson decay constant $F$
receives leading corrections at order~$\mathcal{O}(p^{4})$ in ChPT
 due to the low energy constant $L_5$.
The $\rm{R}\chi{T}$ framework at LO has effectively the
same chiral order. It leads to the same corrections to the decay
constants, when one removes the scalar tadpole term
$\mathcal{L}^{tad}_{S} = {2 c_m} \left\langle S^{} \chi \right\rangle$
and use the hypothesis of resonance saturation~\cite{JJSC:tadpole}.

Let $\phi_8 = \eta_8$ be the octet member, $\phi_0 = \eta_0$ be the
singlet state, and $\lambda_0 \equiv \sqrt{2/3} \; {\rm diag}(1,1,1)$.
Then the pseudoscalar nonet in $\rm{R}\chi{T}$ reads $u \equiv \exp
\left( \frac{\rm i}{\sqrt{2}} \;  \sum_{b=0}^8 \frac{\lambda_b
\phi_b}{\sqrt{2} f_b} \right)$, where $f_b$ had received corrections:
$f_{1,2,3} \neq f_{4,5,6,7} \neq f_8 \neq F$ due to $SU(3)$ flavor breaking and $f_0
\neq f_8 \neq F$ due to nonet symmetry breaking and topological effects of $U(1)$
axial anomaly. The singlet field also obtains an extra contribution to
mass due to $U(1)$ axial anomaly. Translation of the $f_b$
constants into those for physical fields is done in the two-angle mixing
scheme, for a review see~\cite{Feldmann:rev},
\bgea 
\left(
\begin{array}{cc}
\langle 0 | J_{\mu,\,5}^8 (0) | \eta(p) \rangle  & \langle 0 | J_{\mu,\,5}^0 (0) | \eta(p) \rangle  \\
\langle 0 | J_{\mu,\,5}^8 (0) | \eta^\prime(p) \rangle  & \langle 0 | J_{\mu,\,5}^0 (0) | \eta^\prime(p) \rangle
\end{array}
\right)
&=&
{\rm i} \sqrt{2}p_\mu
\left(
\begin{array}{lr}
f_8 \cos \theta_8 & - f_0 \sin \theta_0
\\
f_8 \sin \theta_8 &  f_0 \cos \theta_0
\end{array}
\right)
\equiv
{\rm i} \sqrt{2}p_\mu \hat{f}.
\enea

For the pseudoscalar nonet in physical basis one gets
\bgea
\label{eq:ps-nonet-pre}
u &=&
 \exp \left( \frac{\rm i}{\sqrt{2}} \;  \left[
   \frac{\vec{\pi}\vec{\sigma}}{\sqrt{2} f_\pi}
+\frac{ \lambda_4 K_4 + \lambda_5 K_5 + \lambda_6 K_6 + \lambda_7 K_7}{\sqrt{2} f_K}
+ \frac{1}{\sqrt{2}} \;
\left( \lambda_8\;\;\lambda_0 \right)\! \hat{f}^{-1} \!
\left(
\begin{array}{l}
\eta \\
\eta^\prime
\end{array}
\right)
\right] \right)
\enea
\bgea
\label{eq:etamixing}
&&\left\{
\begin{array}{l}
\eta_8 = \frac{1}{\cos(\theta_8-\theta_0)} \left[ \eta \cos\theta_0  + \eta^\prime \sin\theta_0 \right]
\\
\eta_0 = \frac{1}{\cos(\theta_8-\theta_0)} \left[- \eta \sin\theta_8  + \eta^\prime \cos\theta_8 \right],
\end{array}
\right.
\;\;\;\;\;\;
\left\{
\begin{array}{l}
\eta=\eta_8 \cos\theta_8  - \eta_0 \sin\theta_0
\\
\eta^\prime=\eta_8 \sin\theta_8  + \eta_0 \cos\theta_0 .
\end{array}
\right.
\enea
The mixing angles are determined~\cite{Feldmann:98,Escribano:2005qq} %\cite{Beisert}
from experiment.
We use $\theta_1 = -9.2^\circ \pm 1.7^\circ$ and $\theta_8 = -21.2^\circ \pm 1.6^\circ$~\cite{Feldmann:98},
which correspond to $f_8= (1.26 \pm 0.04) f_\pi$ and $f_0=(1.17 \pm 0.03) f_\pi$.

In order to make the effective lagrangians read simpler one may apply the following notation
\bgea
\label{eq:eta-coefficients}
C_q &\equiv \tfrac{f_\pi}{\sqrt{3} \cos(\theta_8-\theta_0)}
\left( \frac{1}{f_0}\cos\theta_0 - \frac{1}{f_8}\sqrt{2} \sin\theta_8 \right),
\;\;\;\;
\;\;\;\;
C^\prime_q &\equiv \tfrac{f_\pi}{\sqrt{3} \cos(\theta_8-\theta_0)}
\left(\tfrac{1}{f_8} \sqrt{2} \cos\theta_8 + \tfrac{1}{f_0}\sin\theta_0 \right),
\nn
\\
C_s &\equiv \tfrac{f_\pi}{\sqrt{3} \cos(\theta_8-\theta_0)}
\left( \frac{1}{f_0}\sqrt{2}\cos\theta_0 +  \tfrac{1}{f_8}\sin\theta_8 \right),
%\nn
%\\
\;\;\;\; \;\;\;\; C^\prime_s &\equiv \tfrac{f_\pi}{\sqrt{3}
\cos(\theta_8-\theta_0)} \left(\tfrac{1}{f_8}\cos\theta_8 -
\tfrac{1}{f_0}\sqrt{2}  \sin\theta_0 \right) \nn . 
\enea 
Then the pseudoscalar nonet can be written as
 \bgea 
\label{eq:ps-nonet} 
u &\!\!\!\!=& \!\!\!\!
 \exp \left( \frac{\rm i}{\sqrt{2} f_\pi}
\left(
\begin{array}{ccc}
 \frac{\pi^0 + C_{q}\eta + C_{q}^\prime \eta^\prime}{\sqrt{2}} & \pi^+  & \frac{f_\pi}{f_K} {K^+ } \\
 \pi^- & \hspace{-0.018in} \frac{-\pi^0+ C_{q}\eta + C_{q}^\prime \eta^\prime}{\sqrt{2}} & \frac{f_\pi}{f_K} {K^0} \\
 \frac{f_\pi}{f_K} {K^-} & \frac{f_\pi}{f_K} {\bar{K}^0} & \hspace{-0.018in} - C_{s}\eta + C_{s}^\prime \eta^\prime
\end{array}
\right)
\right)
.
\enea

\section{The $\phi(1020)$ radiative decay fits}
\label{sec:Fit}

The dominant decay channels of the scalar mesons are known to be
$\pi^+ \pi^-$, $\pi^0 \pi^0$  for $f_0 (980)$ and $\sigma$ meson,
and $\pi^0 \eta$ for $a_0(980)$ meson.
Much experimental attention has been paid so far to the radiative
decay of the $\phi$ meson: $\phi(1020) \to \gamma a_0 \to \gamma \pi\eta$~\cite{Aloiso02C,KLOEpietaDATA}
and $\phi(1020) \to \gamma f_0\to \gamma \pi\pi$~\cite{KLOEres}.
%(see also~\cite{SNDres,CMD2res,Achasov:2000ym}).
Various models (e.g.~\cite{Ivashyn:2007yy,Harada06}) have been proposed to describe these decays.
In our previous consideration~\cite{Ivashyn:2009yt} we
 performed the separate fits ($\pi^0\pi^0$ and $\pi\eta$) for the mass-parameters of scalar mesons
and the angle $\theta$, while $4c_d c_m = F^2$ and  $c_d=c_m$ relations
of Ref.~\cite{Jamin:2001zq} were imposed. The $\sigma$ meson
contribution was not taken into account and the data points with
$m_{\pi\pi} > 700$~MeV were fitted. 
We ended up with  $\chi^2/{dof} =
2.05\,(\pi^0\pi^0)$, and $\chi^2/{dof}=3.32\,(\pi\eta)$ in a fit to
combined KLOE and Novosibirsk data.

\begin{figure}
\begin{center}
\resizebox{0.49\textwidth}{!}{%
  \includegraphics{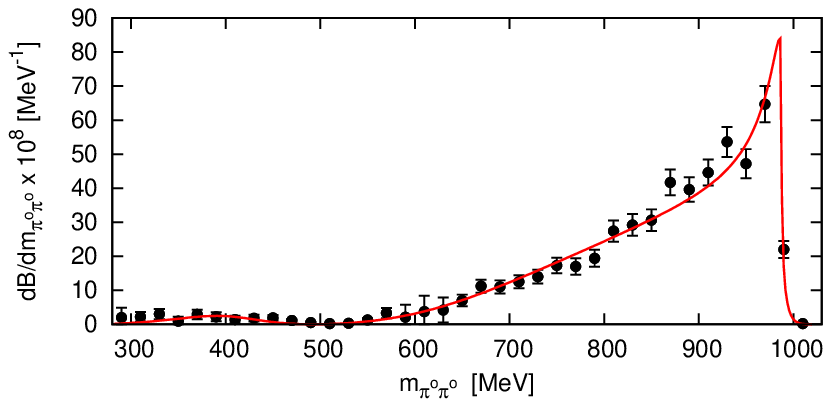}
  }
\resizebox{0.49\textwidth}{!}{%
  \includegraphics{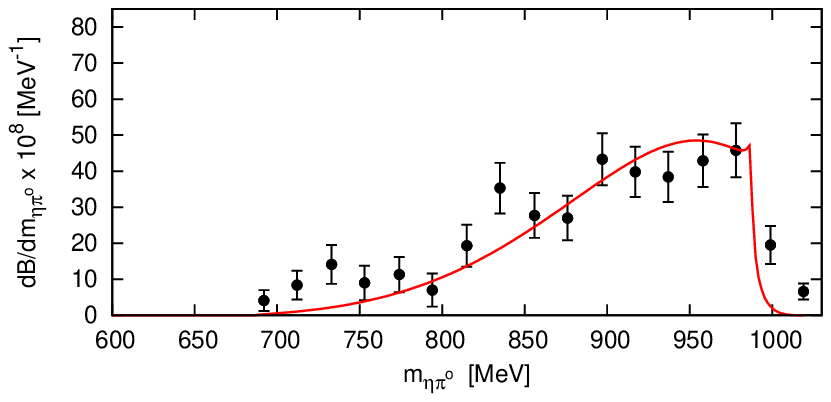}
  }
\end{center}
\caption{
``Best fit'': $\chi^2/d.o.f. (tot)  =  1.47$, $c_d =  93^{+11}_{-5}$, $c_m =   46^{+9}_{-2}$,
$M_{a_0}=1150^{+50}_{-23}$, $M_{f_0}=986.1^{+0.4}_{-0.5}$, $M_{\sigma}=504^{+242}_{-53}$,
 $\theta = 36^o\pm2^o$ (errors from MINOS only).
Data: ~\cite{KLOEpietaDATA,KLOEres}.
 }
\label{fig:1}
\end{figure}

Let $\tilde{M}_S = {M}_S - i/2\; \Gamma$ be the complex pole of the
amplitude. We define the scalar meson propagator (see discussion e.g. in~\cite{Escribano:2002iv}) as 
 \bgea
\label{eq:propagator}
D_{S}^{-1}(p^2)&=& p^2 - \Re\mathrm{e}\left( \tilde{M}_S^2
+ i\, \tilde{M}_S\; \tilde{\Gamma}_{S,\; {tot}}(\tilde{M}_S^2) \right)
+ i\, \sqrt{p^2}\; \tilde{\Gamma}_{S,\; {tot}}(p^2),
 \enea
with the Flatt\'e-modified~\cite{Flatte} widths
 \bgea
\tilde{\Gamma}_{f_0,\;\sigma,\, tot}(p^2) &=&
\tilde{\Gamma}_{f_0,\;\sigma\to \pi\pi}(p^2) + \tilde{\Gamma}_{f_0,\;\sigma\to
K\bar{K}}(p^2),
\\
\nn
\tilde{\Gamma}_{a_0,\, tot}(p^2) &=&
\tilde{\Gamma}_{a_0\to \pi^0\eta}(p^2) + \tilde{\Gamma}_{a_0\to
K\bar{K}}(p^2).
 \enea
These analytic functions of (complex) $p^2$
 (continued below the thresholds through
$\sqrt{{p^2}/{4} - m^2} = i \sqrt{|{p^2}/{4} - m^2|}$) are given
in Appendix~\ref{sec:Propagators}. One should notice that the
exact propagator is of course different from
(\ref{eq:propagator}). Thus
%% we are aware of the fact that
the proper positions of the amplitude poles, $\tilde{M}_S$, do not
coincide with the zeros of $D^{-1}(p^2)$ defined
by~(\ref{eq:propagator}), in other words, $D^{-1}(\tilde{M}_S^2) \ne 0$.
Nevertheless, it is of interest to investigate how the form~(\ref{eq:propagator}) 
reproduces the data. We use the information on
the poles from~\cite{Colangelo,Pennington-Analysis,L3} and perform the
following fits to the $\phi(1020)$ radiative decay spectra of
Refs.~\cite{KLOEpietaDATA,KLOEres}:
%\small
\begin{itemize}
%\begin{enumerate}
 \vspace{-4pt}
 \item[A, B]
                   fixed: $c_d=c_m = F/2 \approx 46.2$~MeV~\cite{Jamin:2001zq};
\hfill (cf. Ref.~\cite{Ivashyn:2009yt})\\
                  fitted (A): $M_{a_0}$ to $\pi\eta$ spectrum, $M_{f_0}$, $\theta$ to $\pi^0\pi^0$ spectrum, $m_{\pi\pi} > 700$~MeV
                            (separate fits);\\
                  fitted (B): $M_{a_0}$ to $\pi\eta$ spectrum, $M_{f_0}$, $M_\sigma$, $\theta$ to $\pi^0\pi^0$ spectrum
                            (separate fits);

 \vspace{-4pt}
 \item[C-E]
                   fixed (C, D): $\tilde{M}_{a_0}=(985\pm10) - i\; (50\pm17)$~\cite{L3}, $\tilde{M}_{f_0}=1001 - i\; 16$~\cite{Pennington-Analysis}; \\
                   fixed (E): $\tilde{M}_{a_0}$~\cite{L3}, $\tilde{M}_{f_0}$~\cite{Pennington-Analysis}, $\tilde{M}_{\sigma} = 441^{+16}_{-8} - i\; 272^{+9}_{-13}$~\cite{Colangelo}\\
                  fitted: $c_d$, $c_m$, $\theta$ to $\pi\eta$ and $\pi^0\pi^0$ spectra (simultaneous fit); $m_{\pi\pi} > 700$~MeV in C;

 \vspace{-4pt}
 \item[E, F]
                  fitted: $M_{a_0}$, $M_{f_0}$, (and $M_{\sigma}$ in F), $c_d$, $c_m$, $\theta$ to $\pi\eta$ and $\pi^0\pi^0$ spectra
                            (simultaneous fit).
%\end{enumerate}
\end{itemize}
%\normalsize
 \vspace{-4pt}
For $\phi(1020) \to \gamma a_0 \to \gamma \pi\eta$ we use only $\eta \to
\gamma \gamma$ data sample of Ref.~\cite{Aloiso02C} and exclude the two
rightmost points from fit. The results are shown in
Table~\ref{tab:fit-results}. The best fit including $\sigma$ meson is
Fit~F, it is illustrated in Fig.~\ref{fig:1}. For Fits~C, D, E we
employed~(\ref{eq:propagator}), while for  Fits~A, B and~F we assumed
$\tilde{M}_s \approx M_s$ in (\ref{eq:propagator}) and employed
 \bgea
D_{S}^{-1}(p^2)&=& p^2 - M_S^2 +  M_S\; \Im\!\mathrm{m}\!\left(
\tilde{\Gamma}_{S,\; {tot}}(M_S^2) \right) + i\, \sqrt{p^2}\;
\tilde{\Gamma}_{S,\; {tot}}(p^2) .
 \enea

\begin{table}
\caption{MINUIT fit results. Shown errors are due to MINOS.
The fixed input values are marked by asterisk.
Couplings and masses are given in MeV.
}
\label{tab:fit-results}
\begin{center}
\small
\vspace{-10pt}
\begin{tabular}{|c|cc|ccc|c|ccc|}
\hline
&&&&&&& \multicolumn{3}{c|}{$\chi^2/{dof}$}
\\
Fit & $c_d$ & $c_m$ & $M_{a_0}$ & $M_{f_0}$ &$M_{\sigma}$  & $\theta$ & $\pi\pi$ & $\pi\eta$ &$tot.$
\\
\hline
A & $46.2^\ast$ & $46.2^\ast$ & $1030\pm7$ & $979.8^{+0.9}_{-0.9}$ & ---  & $21^o\pm1^o$ &$2.95$ & $4.73$ & ---
\\
B & $46.2^\ast$ & $46.2^\ast$ & $1030\pm7$ & $979.0^{+0.8}_{-0.8}$ & $441^{+15}_{-25}$  & $21^o\pm1^o$ & $2.78$ & $4.73$ & ---
\\
\hline
C & $68\pm4$ & $27\pm3$ & $985^\ast$ & $1001^\ast$ &   --- & $22^o\pm2^o$ & $14.25$ & $12.93$ &$12.7$
\\
D & $131^{+14}_{-12}$ & $73^{+7}_{-5}$ & $985^\ast$ & $1001^\ast$ &  ---  & $23^o\pm1^o$ & $6.25$ & $2.48$ &$5.15$
\\
E & $158^{+12}_{-12}$ & $67^{+4}_{-5}$ & $985^\ast$ & $1001^\ast$ & $441^\ast$  & $30^o\pm1^o$ & $4.12$ & $3.30$ &$3.77$
\\
\hline
F & $93^{+11}_{-5}$ & $46^{+9}_{-2}$ & $1150^{+50}_{-23}$ & $986.1^{+0.4}_{-0.5}$ & $504^{+242}_{-53}$  & $36^o\pm2^o$ & $1.32$ & $2.07$ &$ 1.47$
\\
\hline
\end{tabular}
\end{center}
\end{table}

\section{Conclusions}
\label{sec:Discussion}

\vspace{-1pt}
The current results are the good illustration of the model
machinery. There are  important theoretical issues in the scalar
sector of $\rm{R}\chi{T}$. Scalar tadpoles and their relevance for the
corrections to pseudoscalar meson decay constants and problem of
consistent mass splitting for resonances are among them. It was also not
widely known that the account for the $\eta-\eta^\prime$ mixing
at leading order in $\rm{R}\chi{T}$ requires the two-angle scheme in the
singlet-octet basis.

We have performed several fits to the $\phi(1020)$ radiative
decay spectra~\cite{KLOEpietaDATA,KLOEres} in order to fix the
$\rm{R}\chi{T}$ parameters in the scalar sector. It is observed,
that the quality of the fit strongly depends on the form
used for the scalar propagator. We used the information on the
pole positions in order to pin down the masses of scalars in the fit and
concluded that this way seems problematic within the currently
used Flatt\'e-like framework. Our best result is the combined Fit~F,
in which $\sigma$ meson contribution is accounted for. 
It covers the full range of
the invariant masses and has total $\chi^2/d.o.f. = 1.47$.

Appropriate and numerically optimal way to fix the
model parameters is still to be developed. The above
consideration gives a strong motivation for the further improvement of
the model.

\small
\acknowledgments

\vspace{-2pt}
This work benefited from discussions with H.~Czy\.z, R.~Escribano and J.~\!J.~Sanz~Cillero.
We thank M.R.~Pennington for the interest to this work and for suggestions.
A.K. acknowledges support by the INTAS grant 05-1000008-8328.
S.I. was supported by EU-MRTN-CT-2006-035482 (FLAVIAnet),
 partly by Joint NASU-RFFR Project N~38/50-2008 and acknowledges the
hospitality of the Intitut f\"ur Theoretische Teilchenphysik of the Karlsruhe University,
where a part of this work was done.
We warmly thank the Organizers of the Workshop for the very fruitful meeting.

\normalsize
\appendix

\section{Model details for the scalar meson propagators}

\vspace{-2pt}
\label{sec:Propagators} Here we list the reference formulae (see
also~\cite{Ivashyn:2009yt}). The momentum-dependent widths read
 \bgea
\tilde{\Gamma}_{f_0,\sigma\to\pi\pi,KK}(p^2)
& = & \frac{3}{2} \frac{1}{2 p^2} \sqrt{\frac{p^2}{4} -
m_{\pi, K}^2} \; % \times\,
{\frac{G_{f_0,\sigma\pi\pi,KK}^2(p^2)}{4\pi}},
\\
\nn
\tilde{\Gamma}_{a_0\to\pi\eta}(p^2) &=& \frac{1}{2 p^2}
\sqrt{\frac{(p^2+m_\pi^2-m_\eta^2)^2}{4p^2}-m_\pi^2}
\; \frac{G_{a_0\pi \eta}^2(p^2)}{4\pi},
\\
\nn \tilde{\Gamma}_{a_0\to K\bar{K}}(p^2)
&=&2\, \frac{1}{2 p^2} \sqrt{\frac{p^2}{4} -
m_K^2} \; %\times\,
{\frac{G_{a_0KK}^2(p^2)}{4\pi}}.
 \enea
It is assumed that $\sqrt{f(p^2)} = e^{i\; Arg(f(p^2))/2}
\sqrt{|f(p^2)|}$ if necessary (below the two-kaon threshold). In
the $\rm{R}\chi{T}$ formalism the effective couplings for the scalars
are momentum-dependent:
%\noindent
 \bgea
\label{eq:SPP-ffs}
    G_{f_0,\;\sigma KK}(p^2) &\equiv & 1/f_K^2\; \left( \hat{g}_{f_0,\;\sigma KK} (m_K^2 - p^2/2) + g_{f_0,\;\sigma KK} \right), \\
    G_{f_0,\;\sigma \pi \pi}(p^2) &\equiv & 1/f_\pi^2\; \left( \hat{g}_{f_0,\;\sigma \pi \pi} (m_\pi^2 - p^2/2) + g_{f_0,\;\sigma \pi \pi} \right), \nn\\
    G_{a_0 KK}(p^2) &\equiv & 1/f_K^2\; \left( \hat{g}_{a_0KK} (m_K^2 - p^2/2) + g_{a_0KK} \right), \nn\\
    G_{a_0 \pi \eta}(p^2) &\equiv & 1/f_\pi^2\; \left(
     \hat{g}_{a\pi\eta} (m_\eta^2 + m_\pi^2 - p^2)/2 + g_{a\pi\eta} \right).\nn
 \enea
The Lagrangian parameters enter these formulae via

\vspace{-2pt}
\noindent
\parbox[b][150pt][t]{0.5\textwidth}{
\noindent
 \bgea
    g_{aKK}            &=& - \sqrt{2} \, c_m m_K^2, \nn\\
    g_{a\pi\eta}       &=& -2 \sqrt{2}\, C_{q} \,c_m \, m_\pi^2  , \nn\\
    \hat{g}_{aKK}      &=& \sqrt{2} \, c_d, \nn\\
    \hat{g}_{a\pi\eta} &=& 2 \sqrt{2} \, C_{q} \, c_d \nn\\
    g_{f\pi\pi} &=& - 2 \, c_m \, m_\pi^2 (2\, \cos \theta - \sqrt{2} \, \sin \theta)/\sqrt{3},   \nn\\
    g_{fKK} &=& - c_m \, m_K^2(4 \, \cos \theta + \sqrt{2}\, \sin \theta)/\sqrt{3}, \nn\\
    \hat{g}_{f\pi\pi} &=& 2 \,c_d (2 \cos \theta - \sqrt{2}\, \sin \theta)/\sqrt{3}, \nn\\
    \hat{g}_{fKK} &=& c_d(4 \cos \theta + \sqrt{2} \sin \theta)/\sqrt{3}, \nn
 \enea
}
\parbox[b][150pt][t]{0.47\textwidth}{
\bgea
    g_{{\sigma}\pi\pi} &=& - 2 \, c_m \, m_\pi^2 (\sqrt{2}\, \cos \theta + 2 \, \sin \theta)/\sqrt{3},  \nn\\
    g_{{\sigma}KK} &=& - c_m \, m_K^2( - \sqrt{2} \, \cos \theta + 4\, \sin \theta)/\sqrt{3}, \nn\\
    \hat{g}_{{\sigma}\pi\pi} &=& 2 \,c_d (\sqrt{2} \cos \theta + 2\, \sin \theta)/\sqrt{3}, \nn\\
    \hat{g}_{{\sigma}KK} &=& c_d(- \sqrt{2} \, \cos \theta + 4\, \sin \theta)/\sqrt{3}.\nn
 \enea
}


\begin{thebibliography}{99}
\vspace{-5pt}
        \bibitem{EckerNP321}
        G.~Ecker, J.~Gasser, A.~Pich, E.~de~Rafael,
        \emph{The role of resonances in chiral perturbation theory,}
        \emph{Nucl.~Phys.}~\textbf{B~321} (1989) 311.

%         %\cite{Rosell:09:saturation}
%         \bibitem{Rosell:09:saturation}
%         I.~Rosell, P.~Ruiz-Femenia, J.J.~Sanz Cillero,
%         \emph{Resonance saturation of the chiral couplings at NLO in $1/N_C$,}
%         { \tt arXiv:0903.2440 [hep-ph]}.

%         %\cite{Rosell:s-ps-renorm}
%         \bibitem{Rosell:s-ps-renorm}
%         I.~Rosell, P.~Ruiz-Femenia, J.~Portoles,
%         \emph{One-loop renormalization of resonance chiral theory: scalar and pseudoscalar resonances,}
%         \emph{JHEP} \textbf{0512} (2005) 020
%         [{\tt  hep-ph/0510041}].

%         %\cite{Trnka:v-renorm}
%         \bibitem{Trnka:v-renorm}
%         K.~Kampf, J.~Novotn\'y, J.~Trnka,
%
%         { \tt arXiv:0803.1731v1 [hep-ph] }.

        % J.~Trnka,
        % ``Problems with renormalization of effective field theories for resonances,''
        % in {\it Proceedings of Science}, EFT-09, Valencia.

\vspace{-3pt}
        %\cite{Cirigliano:resonances}
        \bibitem{Cirigliano:resonances}
        V.~Cirigliano, G.~Ecker, H.~Neufeld and A.~Pich,
        \emph{Meson resonances, large-$N_c$ and chiral symmetry,}
        \emph{JHEP} \textbf{0306} (2003) 012
        [{\tt hep-ph/0305311}].

%         \bibitem{Bugg:2004xu}
%         D.~V.~Bugg,
%         \emph{Four Sorts of Mesons,}
%         %``Four Sorts of Mesons,''
%         \emph{Phys.\ Rept.\  }{\bf 397} (2004) 257
%         [{\tt hep-ex/0412045}].
%         %%CITATION = PRPLC,397,257;%%

\vspace{-3pt}
        \bibitem{PDG_2008} C. Amsler et al.,
          \emph{The Review of Particle Physics, }
          \emph{Phys.~Lett.}~{\bf B~667} (2008) 1.

\vspace{-2pt}
        %\cite{SCADRON70}
        \bibitem{SCADRON70}
        Proceedings of
        \emph{SCADRON70: Workshop on Scalar Mesons and Related Topics Honoring Michael Scadron's 70th Birthday},
        ed. G.~Rupp {\it et al.},
        \emph{AIP Conf.\ Proc.\  }{\bf 1030} (2008); ISBN 978-7354-0554-7.

%         \bibitem{Pennington2}
%         M.R.~Pennington,
%         \emph{Scalars in the hadron world: the Higgs sector of the strong interaction},
%          \emph{Int.~J.~Mod.~Phys.}~\textbf{A~21} (2006) 747,
%         [{\tt hep-ph/0509265}].

%         %\cite{Hooft:2008we}
%         \bibitem{Hooft:2008we}
%         G.~t'Hooft, G.~Isidori, L.~Maiani, A.~D.~Polosa and V.~Riquer,
%         \emph{A Theory of Scalar Mesons,}
%         %``A Theory of Scalar Mesons,''
%         \emph{Phys.\ Lett.\  } {\bf B 662} (2008) 424
%         [{\tt 0801.2288 [hep-ph]}].
%         %%CITATION = PHLTA,B662,424;%%

\vspace{-1pt}
        %\cite{Black:1998wt}
        \bibitem{Black:1998wt}
        D.~Black, A.~H.~Fariborz, F.~Sannino and J.~Schechter,
        \emph{Putative Light Scalar Nonet,}
        %``Putative Light Scalar Nonet,''
        \emph{Phys.\ Rev.\ } {\bf  D 59} (1999) 074026
        [{\tt hep-ph/9808415}].
        %%CITATION = PHRVA,D59,074026;%%

\vspace{-2pt}
        %\cite{Ivashyn:2007yy}
        \bibitem{Ivashyn:2007yy}
        S.~Ivashyn and A.~Y.~Korchin,
        \emph{Radiative decays with light scalar mesons and singlet-octet mixing in ChPT,}
        %``Radiative decays with light scalar mesons and singlet-octet mixing in
        %ChPT,''
        \emph{Eur.\ Phys.\ J.\ } {\bf  C 54} (2008) 89
        [{\tt 0707.2700 [hep-ph]}].
        %%CITATION = EPHJA,C54,89;%%

\vspace{-2pt}
        \bibitem{Harada06}
        D.~Black, M.~Harada, J.~Schechter,
        \emph{Chiral approach to phi radiative decays,}
        \emph{Phys.~Rev.}~\textbf{D~73} (2006)
        054017.

\vspace{-2pt}
        %\cite{Ivashyn:2008sg}
        \bibitem{Ivashyn:2008sg}
        S.~Ivashyn and A.~Korchin,
        \emph{Radiative decays with $a_0(980)$ and $f_0(980)$ from ChPT at order $p^4$,}
        %``Radiative decays with a0(980) and f0(980) from ChPT at order p^4,''
        \emph{AIP Conf.\ Proc.\  }{\bf 1030} (2008) 123
        [{\tt 0805.4088 [hep-ph]}].
        %%CITATION = APCPC,1030,123;%%

\vspace{-2pt}
        %\cite{Ivashyn:2009yt}
        \bibitem{Ivashyn:2009yt}
        S.~Ivashyn and A.~Korchin,
        \emph{Radiative decays with scalar mesons a0(980) and f0(980) in Resonance Chiral Theory,}
        %``Radiative decays with scalar mesons a0(980) and f0(980) in Resonance Chiral
        %Theory,''
        \emph{Nucl.\ Phys.\ Proc.\ Suppl.\ } {\bf 181-182} (2008) 189
        [{\tt 0901.4045 [hep-ph]}].
        %%CITATION = NUPHZ,181-182,189;%%


%         %\cite{Colangelo:2001df}
%         \bibitem{Colangelo:2001df}
%         G.~Colangelo, J.~Gasser and H.~Leutwyler,
%        \emph{pi pi scattering,}
%         %``pi pi scattering,''
%         \emph{Nucl.\ Phys.\  } {\bf B 603} (2001) 125
%         [{\tt hep-ph/0103088}].
%         %%CITATION = NUPHA,B603,125;%%

\vspace{-2pt}
        \bibitem{Colangelo} I.~Caprini, G.~Colangelo and H.~Leutwyler,
        \emph{Mass and width of the lowest resonance in QCD,}
        \emph{Phys.~Rev.~Lett.} {\bf 96} (2006) 132001.


\vspace{-2pt}
        %\cite{Pennington-Analysis}
        \bibitem{Pennington-Analysis}
        M.R.~Pennington {\it et al.},
        \emph{Amplitude analysis of high statistics results on $\gamma\gamma\to\pi^+\pi^-$
        and the two photon width of isoscalar states,}
        \emph{Eur. Phys. J.} {\bf C 56} (2008) 1.

\vspace{-2pt}
        %\cite{L3}
        \bibitem{L3}
        L3 Collaboration, P.~Achard {\it et al.},
        \emph{$f_1(1285)$ formation in in two-photon collisions at LEP,}
        \emph{Phys.~Lett.} {\bf B 526} (2002) 269.

%         %\cite{DobadoPelaez}
%         \bibitem{DobadoPelaez}
%         A.~Dobado and J.R.~Pel\'aez,
%         \emph{The inverse amplitude method in Chiral Perturbation Theory,}
%        \emph{Phys.\ Rev.\ } {\bf  D 56} (1997) 3057
%         [{\tt  hep-ph/9604416v2}].

\vspace{-2pt}
        %\cite{Oller:2003vf}
        \bibitem{Oller:2003vf}
        J.~A.~Oller,
        \emph{The mixing angle of the lightest scalar nonet,}
        %``The mixing angle of the lightest scalar nonet,''
        \emph{Nucl.\ Phys.\ } {\bf  A 727} (2003) 353.
%        [{\tt hep-ph/0306031}].


%         %\cite{Meissner:tadpole}
%         \bibitem{Meissner:tadpole}
%         V.~Bernard, N.~Kaiser and Ulf G. Meissner,
%         \emph{Chiral perturbation theory in the presence of resonances:
%         application to $\pi\pi$ and $\pi K$ scattering,}
%         \emph{Nucl. Phys.} \textbf{B 364} (1991) 283.

\vspace{-2pt}
        %\cite{JJSC:tadpole}
        \bibitem{JJSC:tadpole}
        J. J. Sanz Cillero,
        \emph{Pion and kaon decay constants: Lattice versus resonance chiral theory,}
        \emph{Phys. Rev.} \textbf{D 70} (2004) 094033
        [{\tt hep-ph/0408080}].

\vspace{-2pt}
        \bibitem{Feldmann:rev}
        T.~Feldmann,
        \emph{Quark structure of pseudoscalar mesons,}
        \emph{Int. J. Mod. Phys.} A \textbf{15} (2000) 159.
%        [{\tt hep-ph/9907491}].

%         \bibitem{Beisert}
%         N.~Beisert, B.~Borasoy,
%         \emph{$\eta-\eta^\prime$ mixing in the $U(3)$ chiral perturbation theory,}
%        {\tt arXiv:hep-ph/0107175}

\vspace{-2pt}
        \bibitem{Feldmann:98}
        T.~Feldmann, P.~Kroll, B.~Stech,
        \emph{Mixing and decay constants of pseudoscalar mesons,}
        \emph{Phys.~Rev.}~\textbf{D~58} (1998) 114006
         [{\tt hep-ph/9802409}].

\vspace{-1pt}
        %\cite{Escribano:2005qq}
        \bibitem{Escribano:2005qq}
        R.~Escribano and J.~M.~Frere,
        \emph{Study of the $\eta-\eta^\prime$ system in the two mixing angle scheme,}
        %``Study of the eta eta' system in the two mixing angle scheme,''
        \emph{JHEP} {\bf 0506} (2005) 029
        [{\tt hep-ph/0501072}].
        %%CITATION = JHEPA,0506,029;%%

\vspace{-2pt}
        \bibitem{Aloiso02C} %a_0
        KLOE Collaboration, A.~Aloisio {\it et al.},
        \emph{Study of the decay $\phi \to \eta\pi^0\gamma$ with the KLOE detector,}
        \emph{Phys.~Lett.}~{\bf B~536} (2002) 209
        [{\tt hep-ex/0204012}].

\vspace{-2pt}
        \bibitem{KLOEpietaDATA} %a_0
%\cite{Bini:2002uk}
%\bibitem{Bini:2002uk}
        KLOE Collaboration, C. Bini {\it et al.},
  %C.~Bini, P.~Gauzzi, S.~Giovanella, D.~Leone and S.~Miscetti,
  \emph{Fit to the invariant mass spectra of the decays $\phi \to \pi^0 \pi^0
  \gamma$ and $\phi \to \eta \pi^0 \gamma$,}
        \emph{KLOE Note}~{\bf 173} (2002).
  %%CITATION = KLOE-NOTE-173;%%


        % \bibitem{KLOE:07}%a_0
        % KLOE Collaboration, F.~Ambrosino {\it et al.},
        % \emph{Study of the radiative decay $\phi \to a_0(980)\gamma$ with the KLOE detector,}
        % {\tt arXiv:0707.4609 [hep-ex]}.

\vspace{-2pt}
        \bibitem{KLOEres} %f_0
        {KLOE Collaboration, A.~Aloisio {\it et al.},
        \emph{Study of the decay $\phi \to \pi^0\pi^0\gamma$ with the KLOE detector,}
        \emph{Phys.~Lett.}~\textbf{B~537} (2002) 21
        [{\tt hep-ex/0204013}].}

        % \bibitem{KLOEres:07} %f_0
        % KLOE Collaboration, F.~Ambrosino {\it et al.},
        % Eur. Phys. J~\textbf{C~49} (2007)  473
        % [{\tt hep-ex/0609009}].

%         \bibitem{SNDres}
%         {SND Collaboration, M.N.~Achasov {\it et al.},
%         \emph{The $\phi \to \eta\pi^0\gamma$ decay},
%         \emph{Phys.~Lett.}~\textbf{B~479} (2000) 53 }
%
%         \bibitem{CMD2res}
%         {CMD-2 Collaboration, R.R.~Akhmetshin {\it et al.},
%         \emph{Study of the $\phi$ decays into $\pi^0\pi^0\gamma$ and $\eta\pi^0\gamma$ final states,}
%         \emph{Phys.~Lett.}~\textbf{B~462} (1999) 380}
%
%         %\cite{Achasov:2000ym}
%         \bibitem{Achasov:2000ym}
%         SND Collaboration, M.~N.~Achasov {\it et al.},
%         \emph{The $\phi(1020) \to \pi^0 \pi^0 \gamma$ decay,}
%         % ``The Phi(1020) --> pi0 pi0 gamma decay,''
%         \emph{Phys.\ Lett.\ } {\bf B 485} (2000) 349
%         [{\tt hep-ex/0005017}]
%         %%CITATION = PHLTA,B485,349;%%

\vspace{-2pt}
        %\cite{Jamin:2001zq}
        \bibitem{Jamin:2001zq}
        M.~Jamin, J.~A.~Oller and A.~Pich,
        \emph{Strangeness-changing scalar form factors,}
        %``Strangeness-changing scalar form factors,''
        \emph{{Nucl.\ Phys.}} {\bf  B 622} (2002)  279
        [{\tt hep-ph/0110193}].
        %%CITATION = NUPHA,B622,279;%%

\vspace{-2pt}
%\cite{Escribano:2002iv}
\bibitem{Escribano:2002iv}
  R.~Escribano, A.~Gallegos, J.~L.~Lucio M, G.~Moreno and J.~Pestieau,
  \emph{On the mass, width and coupling constants of the f0(980),}
  \emph{Eur.\ Phys.\ J.}\  {\bf C 28} (2003) 107  [{\tt arXiv:hep-ph/0204338}].
  %%CITATION = EPHJA,C28,107;%%

\vspace{-2pt}
        %\cite{Flatte}
        \bibitem{Flatte}
        S.M.~Flatt\'e,
  \emph{Coupled - Channel Analysis Of The Pi Eta And K Anti-K Systems Near K Anti-K
  Threshold,}
        \emph{Phys.~Lett.} {\bf B~63} (1976) 224.

% %\cite{Flatte:1976xu}
% %\bibitem{Flatte:1976xu}
%   S.~M.~Flatte,
%   Phys.\ Lett.\  B {\bf 63} (1976) 224.
%   %%CITATION = PHLTA,B63,224;%%


%         \bibitem{Achasov_Ivanchenko}
%         N.N.~Achasov, V.N.~Ivanchenko,
%         \emph{???????????,}
%          \emph{Nucl.~Phys.}~\textbf{B~315} (1989) 465
%
%
% \bibitem{Hanhart:KKMol-phi}
%   Yu.~S.~Kalashnikova, A.~E.~Kudryavtsev, A.~V.~Nefediev, C.~Hanhart and J.~Haidenbauer,
% \emph{The radiative decays $\phi \to \gamma a_0/f_0$ in the molecular model for the scalar mesons,}
%   %``The radiative decays Phi --> gamma a0/f0 in the molecular model for the
%   %scalar mesons,''
%   \emph{Eur.\ Phys.\ J.\ }  {\bf A 24} (2005) 437
%   [{\tt hep-ph/0412340}].
%
% %\cite{Close:1992ay}
% \bibitem{Close:1992ay}
%   F.~E.~Close, N.~Isgur and S.~Kumano,
% \emph{Scalar Mesons in $\phi$ Radiative Decay: their implications for
% spectroscopy and for studies of CP-violation at $\phi$ factories,}
%   %``Scalar Mesons in $\phi$ Radiative Decay: their implications for
%   %spectroscopy and for studies of CP-violation at $\phi$ factories,''
%   \emph{Nucl.\ Phys.\ }  {\bf B 389} (1993) 513
%   [{\tt hep-ph/9301253}].
%   %%CITATION = NUPHA,B389,513;%%
\end{thebibliography}
\end{document}